\let\Xdocument\document
\documentclass{iau}
\let\document\Xdocument

\usepackage{amsmath}
\usepackage{graphicx}
\usepackage{multirow}
\begin{document}

\lefttitle{Lynn D. Matthews}
\righttitle{Mass Loss in Evolved Stars}

\jnlPage{1}{7}
\jnlDoiYr{2023}
\doival{TBD}
\aopheadtitle{Proceedings IAU Symposium}
\editors{T. Hirota,  H. Imai, K. Menten, \& Y. Pihlstr\"om, eds.}

\title{Mass Loss in Evolved Stars}

\author{Lynn D. Matthews$^{1}$}
\affiliation{$^{1}$Massachusetts Institute of Technology Haystack Observatory, 99 Millstone Road, Westford, MA 01886 USA
                 }

\begin{abstract}
Intense mass loss through
cool, low-velocity winds is a defining characteristic of low-to-intermediate
mass stars during the asymptotic giant branch (AGB) evolutionary stage.
Such winds return up $\sim$80\% of the initial stellar mass to the interstellar medium
and play a major role in enriching it
with dust and heavy elements.
A challenge to understanding the physics underlying AGB
mass loss is its
dependence on an interplay between complex and highly dynamic
processes, including pulsations, convective
flows, shocks,  magnetic fields, and opacity
changes resulting from dust and molecule formation. 
I highlight some examples of
recent advances in our understanding of late-stage
stellar mass loss that are emerging
from radio and (sub)millimeter observations, with a particular focus on those
that resolve the surfaces and extended atmospheres of evolved
stars in space, time, and frequency. 

\end{abstract}

\begin{keywords}
stars: AGB -- stars: mass loss -- stars: winds, outflows -- masers
\end{keywords}

\maketitle

\section{Introduction\protect\label{intro}}
Asymptotic giant branch (AGB) represent the final thermonuclear burning stage in the life of low-to-intermediate
mass stars, including stars like the Sun. The AGB marks the second ascent of the red giant branch for these stars,
following the depletion of their core hydrogen supply and the completion of core helium burning. The internal changes to the structure of
the star during the AGB cause the effective temperature to cool to $\sim$2000-3000~K, while to maintain hydrostatic equilibrium,
the star expands to several hundred times its previous size---reaching a diameter of several astronomical units (AU) across
($\sim6\times10^{13}$~cm). At the same
the resulting stellar luminosity increases to $\sim$5000--10,000~$L_{\odot}$. 
AGB stars become unstable to pulsations and typically undergo radial pulsations with periods of order 1 year, accompanied by
significant changes in the visible light output of the star (as high as $\Delta m_{V}\sim$8~mag). A general overview of the properties
of AGB stars can be found in Habing \& Olofsson (2003).

A consequence of the low effective temperatures of AGB stars is that molecules and dust
are able to form and survive in their extended atmospheres. Importantly, the dust that forms helps to drive
copious rates of mass loss (${\dot M}\sim 10^{-8}$ to $10^{-4}M_{\odot}$ yr$^{-1}$) through cool, dense,
  low-velocity winds ($V_{\rm outflow}\sim 10$--20~km s$^{-1}$). These winds are thus over a million times stronger than
  the current solar wind.

The dramatic mass loss that occurs during the AGB evolutionary phase
has implications for wide range of problems in astrophysics.
The details of AGB mass loss (including its duration, as well as the fraction of the initial stellar mass that is shed)
dramatically impact stellar evolutionary tracks, affecting the maximum luminosity a
given star will reach and the type of stellar remnant that  it will ultimately leave behind (e.g., Rosenfield \textit{et al.} 2014;
Kaliari \textit{et al.} 2014). The mass lost by
AGB stars accounts for $\gsim$50\% of the dust and heavy element enrichment in the Galaxy, thus providing
a primary source of raw material for future generations of
star and planets   (Tielens \textit{et al.} 2005; Karakas 2010). And for extragalactic astronomy and cosmology,
accurate prescriptions for AGB mass loss are crucial for stellar population synthesis calculations (e.g., Salaris \textit{et al.} 2014;
Villaume \textit{et al.} 2015), for
understanding dust production and composition in
external galaxies (e.g., Narayanan \textit{et al.} 2021), for interpreting the integrated starlight of distant galaxies (e.g., McGaugh \& Schombert 2014), and for
devising prescriptions of gas recycling and chemical evolution in galaxy models (e.g., Leitner \& Kravtsov 2011; Gan \textit{et al.} 2019).

This article does not attempt a comprehensive review of AGB mass loss (see instead, H\"ofner \& Olofsson 2018; Decin 2021).
Its main focus is to highlight some of the unique insights that can be gained from
observations at cm and (sub)mm wavelengths 
that resolve AGB stars in space, time, and frequency.

\section{Challenges to Understanding AGB Winds and Mass Loss\protect\label{challenges}}
In contrast to luminous hot stars where the winds are driven by atomic line opacity
(e.g., Lamers \& Cassinelli 1999), AGB winds are thought to be primarily dust-driven, with radiation pressure on dust grains transferring momentum to
the gas through absorption and/or scatting, resulting in material being dragged outward to power a quasi-steady
wind. This basic theoretical
framework for AGB winds was established roughly half a century ago (e.g., Wickramasinghe \textit{et al.} 1966; Kwok 1975). However, despite decades of effort, we still lack a complete and fully predictive theory of AGB
mass loss (see  H\"ofner \& Olofsson 2018).

To first order, dust driving appears to work
relatively well for subsets of AGB stars with carbon-rich atmospheres (C/O$>1$), as the carbonaceous grains that are present
tend to have high opacity to
stellar radiation, enabling efficient momentum transfer and wind driving. However, more generally, this model has limitations.
For example, growing empirical evidence suggests that
real AGB winds may often deviate significantly from the idealized picture of
steady, spherical symmetric  outflows (e.g., Nhung \textit{et al.} 2015; Le~Bertre \textit{et al.} 2016; Decin \textit{et al.} 2020).
Furthermore, the majority of AGB stars have oxygen-rich chemistries (C/O$<1$), and the
silicate-rich grains that form in their extended atmospheres
generally have insufficient infrared opacity to drive the winds with the efficiency needed to account for the observed mass-loss
rates. H\"ofner \textit{et al.} (2016) showed that the effects
of photon scattering may help to alleviate this problem. Nonetheless, a persistent conundrum is that grains require sufficiently cool
temperatures ($\sim$1000--1500~K) and low densities to form and survive, but such conditions are typically not reached interior
to $r\sim2-3R_{\star}$ (i.e., $r\sim$6--7~AU) around a typical AGB star.
Thus some additional process is required to transport material from the stellar ``surface'' into the wind launch
region.

It is now widely believed that pulsation and/or convection play key roles in facilitating AGB mass loss (e.g., Willson \& Bowen
1985; H\"ofner 2016; McDonald \textit{et al.} 2018). In broad terms, the interplay between pulsation and convection produces shock waves
in the extended atmosphere, pushing gas outward; dust formation subsequently occurs in the wake of the shock; and finally, radiation
pressure on the resulting grains drags material outward to power the wind (see Figure~2 of H\"ofner \& Olofsson 2018). However, 
the underlying physics is highly complex, and many details are poorly understood and poorly constrained observationally.

\section{Insights from Studies of Large-scale Circumstellar Ejecta\protect\label{CSE}}
For decades, a  primary means of studying AGB mass loss has been through observations of the
spatially extended circumstellar envelopes (CSEs) of chemically enriched gas and dust
that are a ubiquitous feature of these stars. These CSEs may be observed
using a wide variety of tracers, including
molecular line emission, such as CO (Knapp \textit{et al.} 1998; De Beck \textit{et al.} 2010) or other thermal lines
(Patel \textit{et al.} 2011; Claussen \textit{et al.} 2011); 
far-infrared emission from dust (Young \textit{et al.} 1993; Cox \textit{et al.} 2012), and in some
cases,  scattered optical light (Mauron \& Huggins 2006); far-ultraviolet continuum
(Martin \textit{et al.} 2007; Sahai \& Stenger 2023),
or {\mbox{H\,{\sc i}}}
21-cm line emission from atomic hydrogen (G\'erard \& Le~Bertre 2006; Matthews \textit{et al.} 2013).
Historically,  AGB CSEs were typically envisioned and modeled as
spherically symmetric shells, but many of the aforementioned studies show clearly that
CSE morphologies can be extraordinarily diverse. Depending on the age of the central star, its mass-loss rate, and
the particular observational tracer, the observed extent
of the CSE can range from tens of thousands of AU to a parsec or more, and properties of the CSE can be dramatically shaped by the
presence of
(sub)stellar companions (Maercker \textit{et al.} 2012; Aydi \& Mohamed 2022) or the star's motion through the surrounding interstellar medium
(e.g., Cox \textit{et al.} 2012;
Martin \textit{et al.} 2007; Villaver \textit{et al.} 2012; Matthews \textit{et al.} 2013).

Global studies of CSEs supply a wide array of fundamental information on the mass-loss properties
of evolved stars, including measurements of the mass-loss rate and outflow
speed. In addition, they can provide clues on the nature of the central star (age, temperature, initial
mass), the timescale of the mass-loss history, and the mass-loss geometry (spherical, bipolar, etc.).
Despite the long history of studies of AGB CSEs, observations using the latest generation of radio telescopes continue
to yield new insights and surprises. 
One recent example is the ATOMIUM project\footnote{\url{https://fys.kuleuven.be/ster/research-projects/aerosol/atomium}}, an Atacama
Large Millimeter/submillimeter Array (ALMA) Large Project
that targeted a sample of
AGB stars and red supergiants in the 214--270~GHz range with the goal of obtaining a better understanding of the chemical
and physical processes that govern red giant winds (Decin \textit{et al.} 2020; Gottlieb \textit{et al.} 2022).
Results to date show that asphericity appears to be the norm among AGB ejecta and that
there is a correlation between the morphology of AGB ejecta and the current mean mass-loss rate. This program has also added to growing
evidence that long-period
companions ($P>$1~yr) commonly play a role in shaping CSEs, and that a common mechanism controls the wind morphology of both AGB stars
and planetary nebulae (Decin \textit{et al.} 2020).

Another Large ALMA Project aimed at studying AGB ejecta is DEATHSTAR\footnote{\url{https://www.astro.uu.se/deathstar/index.html}},
which has used the ALMA Compact Array to obtained spatially resolved CO measurements and line profiles for a sample of $\sim$70 chemically
diverse AGB stars. Results to date show that large-scale asymmetries and complex velocity profiles are common. Future radiative transfer modeling
is underway to determine accurate mass-loss rates and temperature distributions of the gas for the sample
(Ramstedt \textit{et al.} 2020; Andriantsaralaza \textit{et al.} 2021).
Meanwhile, the NESS\footnote{\url{https://evolvedstars.space}} program has been conducting a volume-limited survey of $\sim$850
evolved stars in CO and in the sub-mm continuum using the APEX and JCMT telescopes,
with the goal of measuring outflow parameters, gas-to-dust ratios, and other information critical for
characterizing the mass-loss histories of a large sample of stars (Sciclula \textit{et al.} 2022). It is worth emphasizing that single-dish projects like
NESS remain a valuable complement to interferometric surveys such as ATOMIUM and
DEATHSTAR, owing to their ability to target larger samples of stars and to
characterize spatially extended and diffuse molecular emission in CSEs which can be resolved out in interferometric measurements.

\section{Advances in Atmospheric Modeling of AGB Stars\protect\label{mods}}
The complex physics of AGB star atmospheres makes modeling them both challenging and computationally expensive. Approximations of local thermodynamic equilibrium (LTE)
break down in the dynamic, time-varying conditions of AGB atmospheres, and a wide range of physics needs to be included (pulsation, convection,
dust formation, etc.) to produce meaningful results.
Furthermore, because of the enormous spatial extents of AGB star atmospheres and outflows, the
relevant spatial scales required in the model can span many orders of magnitude, ranging from the scales of shock regions
and sub-surface convective cells ($\ll R_{\star}$) to scales of $>1000R_{\star}$ ($>10^{16}$~cm) as required
to fully trace the evolution of temperature, density, and composition of the wind.
Until recently, these challenges have often meant relying on 1D models with simplified physics (e.g., Ireland \textit{et al.} 2011;
Liljegren \textit{et al.} 2018). While instructive for some applications, these models have important limitations. For example,
since the effects of convection are inherently 3D, the result is a blurring of the distinction between pulsation and convective
processes in 1D models (see Freytag \& H\"ofner 2023).
Fortunately, computational advances have begun enabling sophisticated new 3D
radiation-hydrodynamical models that are able to overcome such limitations 
by incorporating a wide range of relevant physics, including radiative transfer, frequency-dependent opacities,
pulsation, convection, shocks, dust formation, grain growth and evaporation, and wind acceleration
(e.g., Freytag \textit{et al.} 2017; Freytag \& H\"ofner 2023).

\begin{figure}[t]
  \begin{center}
    \includegraphics[scale=0.6]{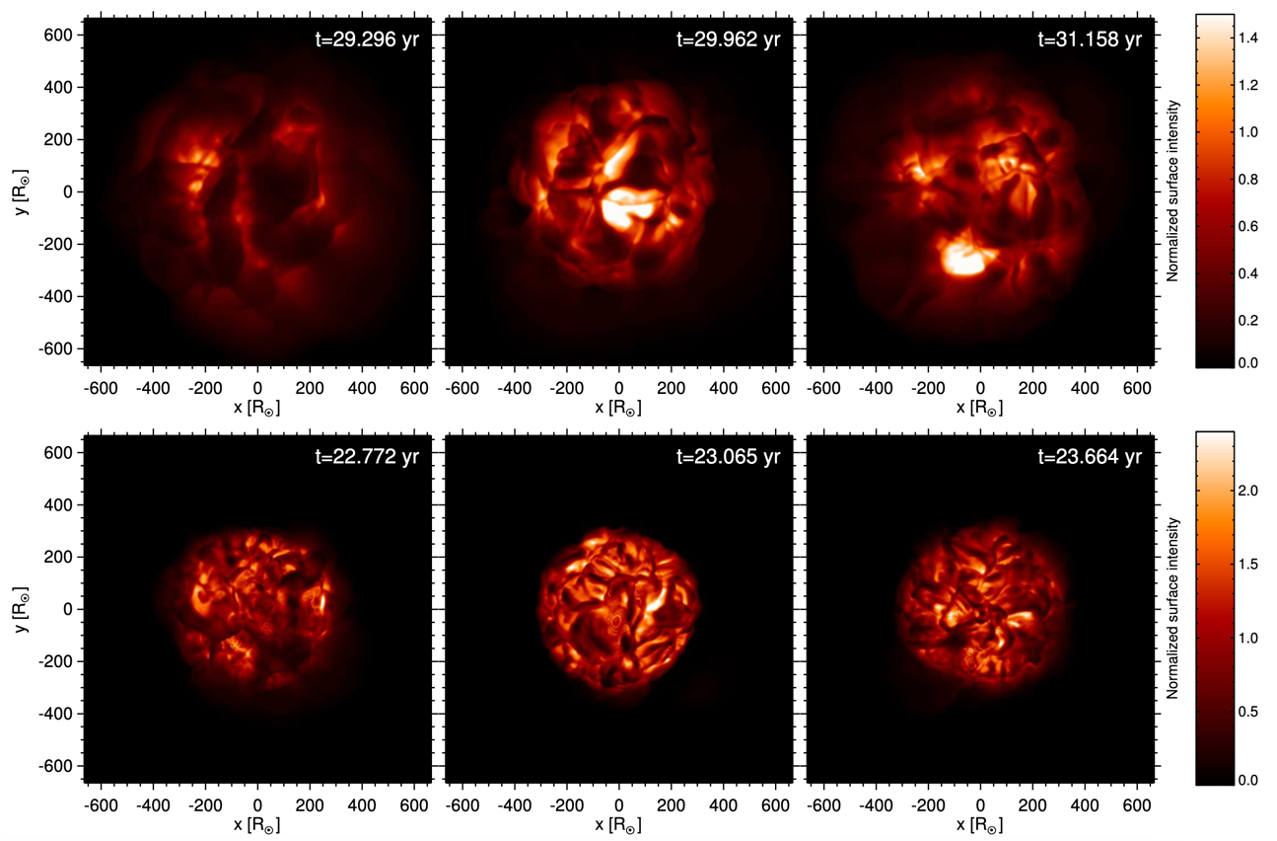}
    \caption{Time sequences of bolometric surface intensity for AGB stars from the 3D CO5BOLD hydrodynamic models of
      Freytag \& H\"ofner (2023). A 1.0~$M_{\odot}$ model is shown in the top row and a 1.5~$M_{\odot}$ model in the bottom row.
      Snapshots are spaced by 8 and 14 months (top) and 3.5 and 7 months (bottom), respectively, from the starting frame.
    The size of each box is $\sim$5.6~AU across.}
  \label{fig:freytag}
  \end{center}
\end{figure}

Figure~\ref{fig:freytag} shows two time sequences of images of bolometric surface intensity from hydrodynamic
simulations of AGB stars from
Freytag \& H\"ofner (2023). The frames
separated by a few months. Both models have a  similar luminosity, but the 1$M_{\odot}$ model (top) exhibits a lower surface gravity, a more
extended atmosphere, and more efficient dust formation, while the 1.5$M_{\odot}$ model (bottom) displays a smaller radius,
a better defined surface,
and smaller, more granular surface features. While we may not yet have a fully predictive theory of stellar mass loss,
models such as these are now giving us incredibly detailed predictions that can be confronted with observations.

\section{Zooming into the Action\protect\label{zoomin}}
Studies of the large-scale CSEs (Section~\ref{CSE})
remain an invaluable tool for characterizing AGB mass loss. However, directly confronting the types of highly detailed
models described above, and
solving many of the outstanding puzzles related to the launch and geometry of AGB winds and their relationship to stellar
pulsations, shocks, convection, and other dynamic phenomena, demands additional
types of observations. In particular, there is a need for observations that:
\vspace{-0.2cm}
\begin{enumerate}
\item {\it Spatially} resolve the stellar atmosphere on relevant physical
  scales (i.e., $r\lsim$10~AU, or even $r\ll R_{\star}$) to probe
  the photosphere, surrounding molecular layers, and the dust formation zone.

\item {\it Temporally} resolve processes on relevant dynamical timescales (which for AGB stars can range from days to years to decades).

\item {\it Spectrophotometrically} distinguish different layers of the atmosphere to trace changes in physical conditions
  and chemistry.

  \item Directly measure gas motions.

\end{enumerate}

\noindent
Fortunately, observations at radio (cm through sub-mm) wavelengths using the latest
generation of radio telescopes
provide a variety of means to achieve  these objectives.
The remainder of this review highlights some examples of recent progress in
our understanding of
AGB mass-loss through radio observations of both molecular masers and thermal continuum emission that
resolve the dynamic behavior of AGB  atmospheres in space and/or time.
I close by briefly highlighting the prospects of future observational facilities for
additional progress in these areas.

\section{Molecular Masers as a Tool for Understanding AGB Mass Loss}
The first detection of molecular maser emission associated with the extended atmosphere of an evolved star was
made by Wilson \& Barrett (1968), who
detected masing OH lines at 1612, 1665, and 1667~MHz from the red supergiant NML~Cyg using the Green Bank 140-ft telescope.
This was followed
a few years later by the detection of H$_{2}$O (at 22.2~GHz) and SiO (at 43.1~GHz) masers, respectively, in other O-rich red giants
(Sullivan 1973; Thaddeus {\it et al.} 1974). Today, molecular masers from various transitions and isotopologues of OH, H$_{2}$O,
and SiO have been detected in hundreds of O-rich Galactic AGB stars and red supergiants (e.g., Engels \& Lewis 1996; Pardo \textit{et al.} 1998;
Kim \textit{et al.} 2010; Rizzo \textit{et al.} 2021), providing a unique resource
for probing the gas dynamics and physical conditions of their
atmospheres. While C-rich AGB stars generally do not give rise to masers from O-bearing species, they may show
maser activity from other molecules, including HCN or SiS (Henkel 1983; Omont \textit{et al.} 1989; see also Section~6.3 below).

More general overviews
of the topic of stellar masers can be found in
e.g., Humphreys \& Gray (2004); Kemball (2007); Colomer (2008); Gray 2012; and Richards (2012).
Here I highlight just a few examples of recent results that illustrate the role 
of maser studies for addressing the objectives outlined in Section~\ref{zoomin}.
I focus primarily on SiO masers, which tend to arise inside
the wind launch region of AGB stars and close to the dust formation zone. In contrast, H$_{2}$O and OH masers
tend to arise at successively larger radii ($\gsim 10^{14}$~cm and $\gsim 10^{15}$~cm, respectively), beyond
the wind launch zone (e.g., Dickinson 1978).
Different transitions and
isotopologues of SiO are further segregated according to the specific combinations of temperature and
density that are necessary to produce masing in each respective line.
In addition to their favorable location in the atmosphere, the compact sizes and high brightness temperatures
(often $>10^{6}$~K)
of SiO masing regions provide the advantage of enabling observations of the masers with extraordinarily high angular resolution ($<$1~mas)
using very long baseline interferometry (VLBI) techniques. For
example, the longest baseline of the Very Long Baseline Array (VLBA) of $\sim$8600~km gives
an angular resolution of $\sim$0.5~mas at 43~GHz, corresponding to a spatial resolution of $\sim$0.1~AU ($\sim0.05R_{\star}$) for a star
at 200~pc.

\subsection{Spatial Distributions}
Thanks to VLBI studies of SiO masers in
a number of AGB stars that have been undertaken
since the 1990s, it is now well established that SiO masers in AGB stars
are typically found to lie (in projection) in ring-like or partial
ring-like structures, with a mean radius of roughly twice that of the stellar
photosphere (e.g., Diamond \textit{et al.} 1994; Cotton \textit{et al.} 2006;
Imai \textit{et al.} 2010). Evidence for spatial segregation is observed between different SiO transitions and isotopologues,
allowing them to be used as probes of changes in physical conditions and gas kinematics over scales $\ll R_{\star}$ (e.g., Desmurs \textit{et al.} 2000;
Wittkowski et al. 2007). This information also provides important constraints on maser pumping models (e.g., Humphreys \textit{et al.} 2002;
Gray \textit{et al.} 2009). 
Unfortunately, a persistent challenge has been that owing to bandwidth limitations of previous
generations of instruments, it was generally not possible to
observe different transitions strictly simultaneously. This, coupled with the typical use of self-calibration procedures (which can
erase absolute astrometric information), meant that there has historically been uncertainty regarding 
the astrometric alignment between different transitions, as well as their locations
relative to the central star. While approximate methods can be used to align the different measurements (e.g., Desmurs \textit{et al.} 2000),
in many cases,
lingering uncertainties can be as high as several mas and potentially
result in  ambiguities in the interpretation (see, e.g., Soria-Ruiz \textit{et al.} 2004).

One example of important progress in overcoming this challenge
has been achieved 
using the Korean VLBI Network (KVN) Multi-Band System (Han \textit{et al.} 2008) together with
the so-called frequency phase transfer (FPT) technique (Dodson \textit{et al.} 2014).
This approach has enabled simultaneous observations of up to five SiO and H$_{2}$O transitions
in several evolved stars (e.g., Dodson \textit{et al.} 2014; Yoon \textit{et al.} 2018; Kim \textit{et al.} 2018).
Currently, maximum KVN baselines are $\sim$450~km. However, future
extension of this technique to longer baselines would be highly desirable to achieve even finer resolution
of individual maser-emitting clumps and to enable improved astrometric precision for following their proper motions over time.

\subsection{Magnetic Field Measurements\protect\label{Bfield}}
Another type of investigation that is possible through observations of stellar masers, including SiO masers, is the study of
polarization and magnetic fields in circumstellar environments (see the overview by
Vlemmings 2012). For example, in a VLBA study of SiO masers in the OH/IR star OH~44.8-2.3,
Amiri \textit{et al.} (2012) measured linear polarization of up to 100\% in individual maser clumps, enabling them to map out the magnetic
field vectors surrounding the star. For the brightest maser clump they also
found evidence of circular polarization, enabling estimation of the magnetic field strength (1.5$\pm$0.3~G). Intriguingly,
both the distribution of the SiO maser clumps and the orientation of the magnetic field vectors surrounding
this star point to a preferred
outflow direction for the stellar wind, hinting that a dipole magnetic field may play a role in shaping and defining the outflow. These
findings in support of a non-spherically symmetric outflow
complement  other recent findings pointing to similar trends based on the study of molecular line emission in AGB stars
on larger scales 
(e.g., Decin \textit{et al.} 2020; Hoai \textit{et al.} 2022b; Winters \textit{et al.} 2022).

\subsection{Masers in Carbon-rich AGB Stars\protect\label{sec:carbonmas}}
As noted above, SiO masers are generally absent in AGB stars with carbon-rich chemistries. However, masing action in C-type
stars has been observed in a few other species, including HCN
(e.g., Omont \textit{et al.} 1989; Izumiura \textit{et al.} 1987; Bieging 2001; Menten \textit{et al.} 2018).
HCN is the most common molecule in the atmospheres of C stars
after H$_{2}$ and CO, although its masing properties have
been relatively little studied to date compared with SiO masers in O-rich stars.

\begin{figure}[t]
  \begin{center}
    \includegraphics[scale=0.4]{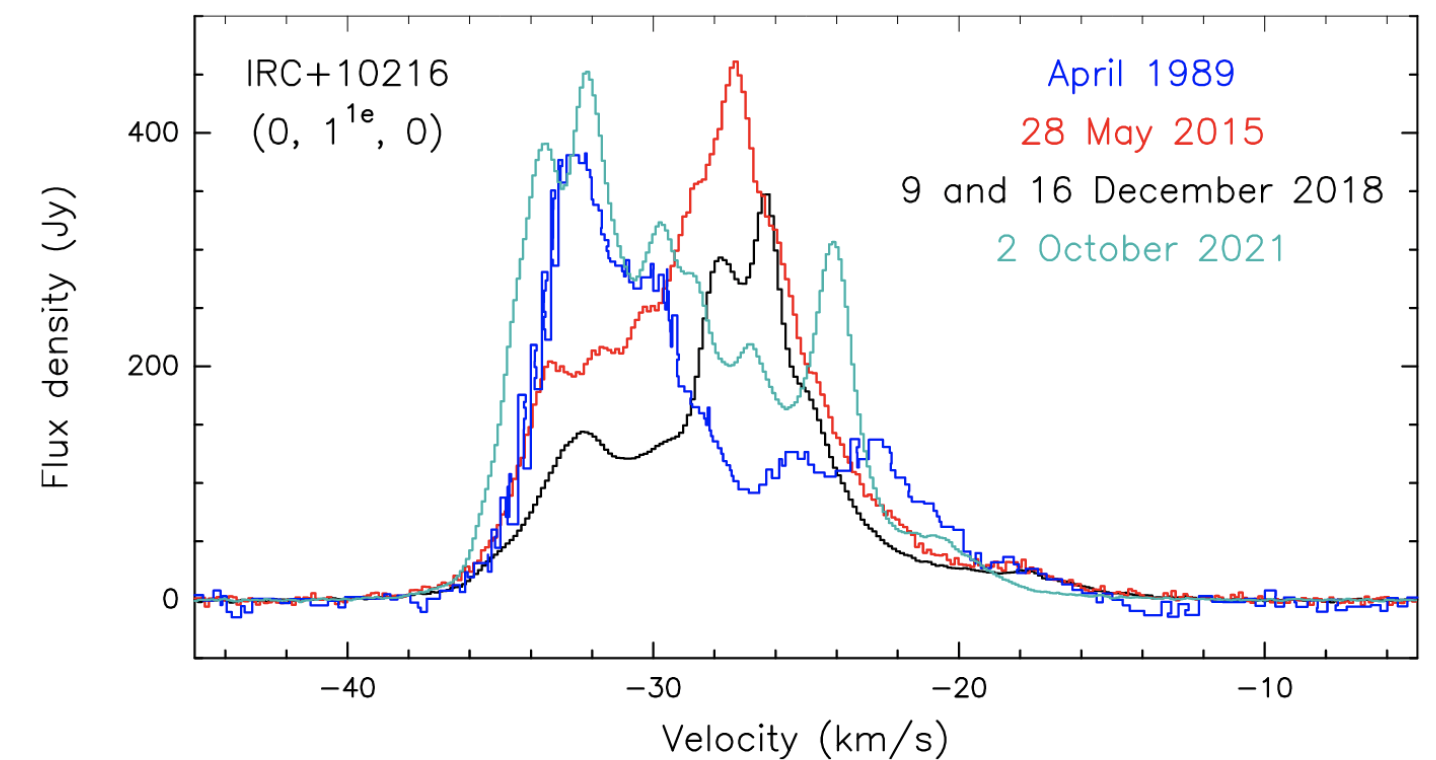}
    \caption{Observed variability of the HCN  (0,1$^{1\rm e}$,0) $J=2-1$ maser at 177.2~GHz in the carbon star IRC+10216. The different
      colored lines show the results from different observing dates. From
      Jeste \textit{et al.} (2022).}
  \label{fig:carbon}
  \end{center}
\end{figure}

If we wish to study the inner regions of the CSEs of C-type stars
using HCN masers (in a manner analogous to what is possible using SiO masers in O-type AGB stars)
it is helpful to target higher $J$, vibrationally-excited states of HCN,
where the opacity is lower. Studies of these transitions have been limited until now, owing to a dearth of observational
facilities equipped with receivers covering the necessary frequency range
($\nu>$176~GHz). However, recently this has begun to change. For example,
Jeste \textit{et al.} (2022) surveyed a sample of 13
C-type stars using the APEX telescope and bands centered at 180~GHz, 230~GHz, and 345~GHz, respectively, providing access to 26 different
HCN transitions. Masing was observed in several different
transitions, including the HCN  (0,1$^{1\rm e}$,0)  $J=2-1$ ($\nu_{0}$=177.2~GHz) line, which was detected
in 11 targets, suggesting that it is a common feature of carbon stars.
Furthermore, the observed
  velocity extents of theses masers indicate that the lines are originating in the acceleration zone where dust is forming, implying
  they have the potential to serve as an important
  new diagnostic tool for the study of wind launching in C-type AGB stars. For stars with multi-epoch observations, clear
  changes in the line profile were seen with time  (e.g., Figure~\ref{fig:carbon}), including in some cases, over
  the course of only a few days.

\subsection{The Time Domain\protect\label{timedomain}}
\subsubsection{Global Measurements}
Adding a temporal dimension of maser studies significantly expands what we can learn about the time-varying atmospheres
of AGB stars compared with single-epoch observations. As noted in Section~\ref{intro},
radial pulsations are a defining characteristic of AGB stars, and these commonly have periods of order 1 year. These are accompanied by
changes in the visible light output of the star by factors of up to a thousandfold (e.g., Reid \& Goldston 2002 and references therein).
In the case of SiO masers, it has long been recognized that variations in the SiO line profiles are correlated with the pulsation
cycle. For example, for Mira-type variables, the study of Pardo \textit{et al.} (2004) found a correlation between the integrated intensity of
the SiO $v$=1, $J=1-0$ line at
43.1~GHz and both the infrared and optical light curves. The masers tend to vary in phase with the infrared, but lag the optical light curve phase
by $\sim$0.05 to 0.2. The authors cited this as evidence that the SiO masers must be radiatively pumped. However, secular
variations of the SiO masers not obviously linked with the pulsation cycle were also seen.

While there have been numerous time-domain studies of SiO and other stellar masers over the past few decades, our ability to
fully exploit and interpret the
results for understanding AGB star atmospheres and mass-loss (as well as the underlying maser physics)
has been hampered by several limitations of these studies.
Among these are: (i) limited instrumental bandwidths (which have precluded simultaneously monitoring multiple maser transitions); (ii)
limited spectral resolution (which obscures the complex velocity structure of the line and may make it impossible to discern subtle changes with time);
(iii) limited signal-to-noise ratios (preventing the detection of weaker lines); (iv)
sample selection biases [e.g., exclusion of semi-irregular and irregular (non-Mira-type) variables]; (v)
and monitoring programs
which are either short-lived (a few years or less) and/or sparsely sampled (observing cadences of $\ge$1 month), thus producing
observations which are
unable to sample all relevant dynamical timescales for the stars. Fortunately, recent progress has been made in nearly
all of these areas.

One example of the power of simultaneously monitoring multiple lines with a wide frequency band is the
study by Rizzo \textit{et al.} (2021), which surveyed  67 O-rich AGB stars and red supergiants
between $\lambda$7~mm and $\lambda$1~mm, targeting SiO rotational
transitions between $J=1-0$ and $J=5-4$, vibrational numbers $v$=0 to 6, and 3 different isotopologues
($^{28}$SiO, $^{29}$SiO, and $^{30}$SiO).
This study resulted in the detection of several new  SiO lines in many of the targets, thus revealing the fascinating complexity of their
multi-wavelength SiO spectra. Among these was first detection of an SiO $v$=6 line.
Additionally, dramatic variations in the line profiles of some targets were seen on timescales as short as $\sim$2 weeks.

Evidence for SiO maser variability over even shorter timescales was reported by
G\'omez-Garrido \textit{et al.} (2020).
These authors performed daily monitoring of a sample of 6 stars and found evidence of rapid ($\sim$1 day) intensity variations
of $\sim$10--25\% in multiple SiO lines in two semi-regular variables (RX~Boo and RT~Vir).
Similar variations were not seen in the Mira-type
variables in the sample. The authors postulated that the semi-regular variables may have intrinsically smaller maser-emitting clumps
and more chaotic shock behaviors in their atmospheres. However, high-cadence VLBI monitoring observations of
semi-regular variables will be
needed to test these ideas and improve our understanding of these phenomena.

\subsubsection{Spatially and Temporally Resolved Imaging Spectroscopy\protect\label{masermovie}}
As described above, multi-epoch maser measurements provide important insights into the time-varying behavior of AGB
star atmospheres.  When this is combined with spatially resolved measurements (particularly with VLBI resolution), our ability
to interpret the results in a physically meaningful way is significantly enhanced (e.g., Richards \textit{et al.} 1999;
Gray \& Humphreys 2000; Phillips \textit{et al.} 2001; Wittkowski \textit{et al.} 2007).

\begin{figure}[t]
  \begin{center}
    \includegraphics[scale=0.4]{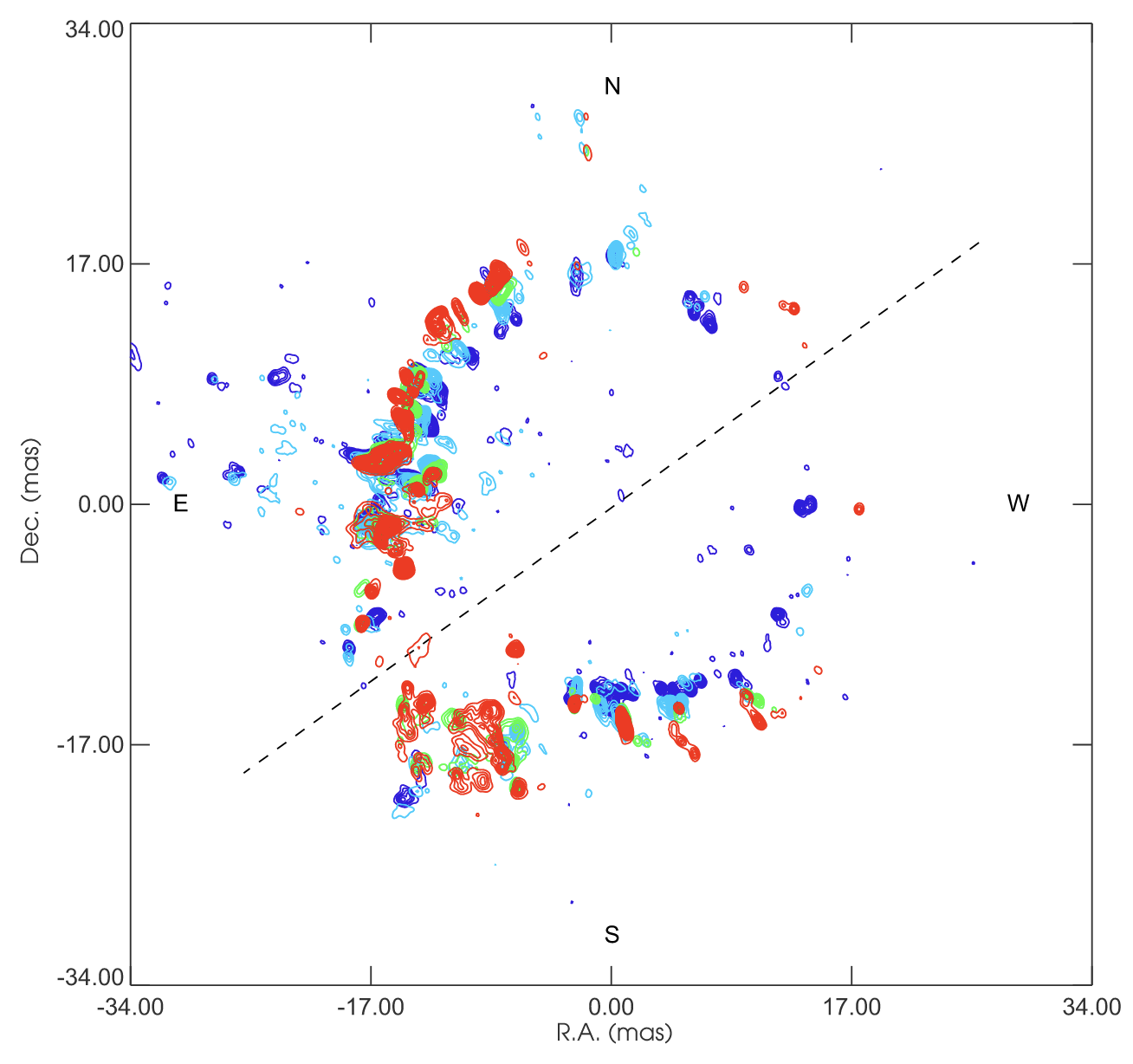}
    \caption{Contour maps of the velocity-integrated
      SiO $v$=1, $J=1-0$ maser emission in TX~Cam, as observed during four different epochs over the span of $\sim$7 months.
      Data from each epoch are indicated by
      a different color. Based on the proper motions of the maser spots between epochs, it is apparent that the expansion velocity
      is higher along the SE-NW axis (dashed line)
      compared with the NE-SW axis, indicating a bipolar geometry. From Gonidakis \textit{et al.} (2013). }
  \label{fig:txcam}
  \end{center}
\end{figure}

Undoubtedly one the most spectacular examples of the power of spatially resolved maser monitoring observations for the study
of evolved stars is the 78-epoch
study of the SiO $v$=1, $J=1-0$ masers in the Mira variable TX~Cam undertaken by Gonidakis \textit{et al.} (2013). Using the VLBA,
these authors observed
the star on a $\sim$2--4 week cadence over the course of nearly 5 years, resulting in a dramatic ``movie'' of the star's evolving atmosphere.
This study confirmed that the proper motions of SiO maser clumps can be used to
trace gas motions close to the stellar photosphere, revealing
both the expansion and infall of gas. The width and boundary of the SiO maser ``ring'' in TX~Cam (actually a shell seen in projection)
was found to vary with stellar pulsation phase,
and evidence for the creation of shocks with velocity of $\sim$7~km s$^{-1}$ was observed during each pulsation cycle. These shocks in turn
affected the intensity and variability of the masers. Importantly, the TX~Cam observations showed no evidence of strong shocks
($>$10 km s$^{-1}$), in agreement with past analysis of radio continuum light curves of other AGB stars (Reid \& Menten 2007; see also
Section~\ref{radiophot}. This supports a model where stronger
shocks are damped by the time they reach $r\sim2R_{\star}$. Additionally, the distribution and velocity structure of the masers are strongly
suggestive of a bipolar outflow (Figure~\ref{fig:txcam}), adding to evidence that such geometries are in fact commonplace for AGB mass loss (see
also Sections~\ref{Bfield}
\& \ref{CSE}). Although the TX~Cam movie is now a decade old, it is worth highlighting again here to emphasize the incredible scientific richness of
data sets of this kind for understanding the physics of AGB star atmospheres 
and to underscore the importance of undertaking similar observations in the future for additional AGB stars spanning a range of properties.

\section{Studies of Radio Photospheres\protect\label{radiophot}}
Despite the many advantages of masers for probing the atmospheric properties and mass-loss physics of evolved stars, such
studies suffer
from certain limitations. For example, maser emission is not observed in all AGB stars, and
for some AGB stars, the maser
emission may become at times too weak to detect. In addition, the interpretation of changes
in the maser emission over time can be challenging in cases where the spatial distribution of the
maser clumps is not spatially resolved,
or where only a single
observing epoch is available. In such instances, it can be difficult to distinguish
changes resulting from varying physical conditions (e.g., changes in temperature or density)
from changes caused by motions of the maser-emitting gas.
Fortunately, recent advances in other observational techniques can help to
provide complementary information, including observations of thermal
continuum emission from the atmospheric region known as the {\it radio photosphere}.

The existence of so-called radio photospheres in AGB stars was first established by Reid \& Menten (1997). These authors examined
a sample of nearby AGB stars and found that
the flux densities at cm through far-infrared wavelengths were systematically higher than predicted
from a simple blackbody model based on the known stellar effective temperatures. This led Reid \& Menten  to postulate that the stars must have
an optically thick layer (i.e., a radio photosphere) lying at $r\sim2R_{\star}$.
They developed a model for the radio photosphere in which the
opacity arises primarily from interactions of free electrons with neutral H and H$_{2}$. For a typical O-rich AGB star, the $\tau$=1 surface
of the radio photosphere lies at $r\sim$2--3~AU and the spectral index of the emission is
slightly shallower than a blackbody ($\alpha\approx$1.86).

\begin{figure}[t]
  \begin{center}
    \includegraphics[scale=0.55]{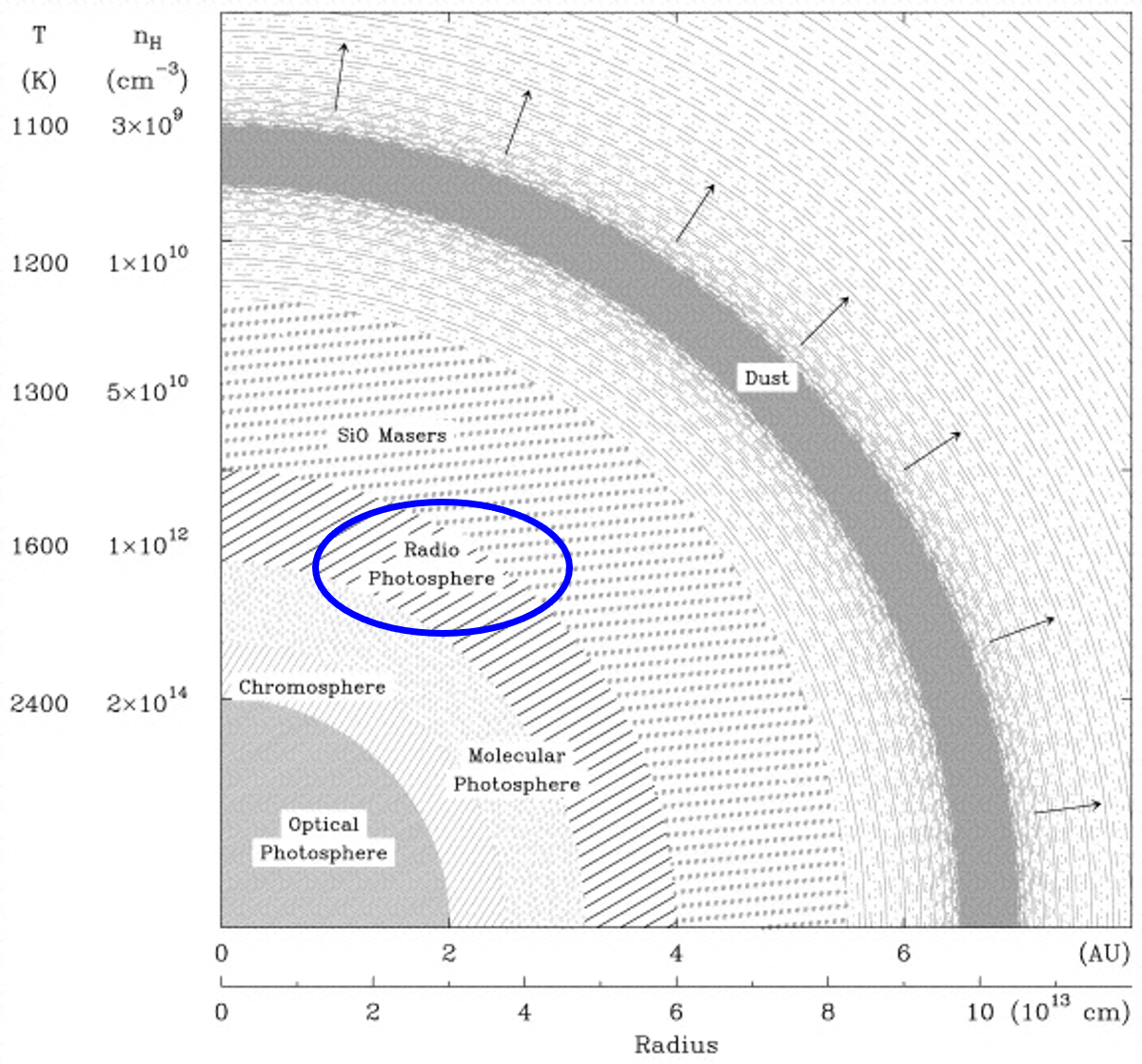}
    \caption{Schematic cross-section of the atmospheric layers in a typical O-rich AGB star.
      A radio photosphere lies at $\sim2R_{\star}$, just interior to the dust formation zone and wind launch region.
      The properties of the radio photosphere are therefore susceptible to the underlying physical processes that
      help to launch the wind, including pulsation, convective flows, and shocks. The radio photosphere is also adjacent to
      the region that gives rise to SiO maser emission in many AGB stars. Adapted from Menten \& Reid (1997). }
  \label{fig:reid}
  \end{center}
\end{figure}

Figure~\ref{fig:reid} shows schematically where the radio photosphere lies in relation to the other atmospheric layers in a typical O-rich AGB star. Crucially,
the radio photosphere resides in the
zone between the classical photosphere and the wind launch region. A consequence is that the properties of the radio
photosphere will be impacted by the shocks, pulsation, convection, and other key physical processes that are believed to be responsible for
helping to transport material into the wind launch region at  $r\sim10R_{\star}$
(Reid \& Menten 1997; Gray \textit{et al.} 2009; see also Section~\ref{challenges}).

The emission from radio photospheres is thermal, and its brightness temperature is too low to be studied at ultra-high angular
resolution using VLBI techniques. However, the radio photospheres of nearby AGB stars ($d\lsim$200~pc) can be resolved with the longest
baseline configurations of the Very Large Array (VLA) and ALMA.
Using $\lambda$7~mm observations obtained with the legacy VLA,
Reid \& Menten (2007) and Menten \textit{et al.} (2012) produced the first spatially resolved images of
the radio photospheres of 4 nearby AGB stars (the O-rich stars Mira, W~Hya, R~Leo, and the carbon star IRC+10216, respectively).
The three O-rich stars also exhibit SiO masers, and Reid \& Menten (2007) used
simultaneous observations of the $\lambda$7~mm continuum and the SiO maser emission to establish unambiguously for the first time
that the SiO masers are distributed in a shell exactly centered on the stellar photosphere.

Another
key finding to emerge from the above studies was that some of the radio photospheres showed clear evidence for deviation from spherical
symmetry. However, with only a single measurement epoch, it was impossible to discern
whether these shapes were static or time-varying.
Taking advantage of the order-of-magnitude boost in continuum sensitivity of the upgraded Karl G. Jansky VLA, Matthews \textit{et al.} (2015, 2018)
reobserved the stars studied by Reid \& Menten (2007) and Menten \textit{et al.} (2012). The resulting observations confirmed
that asymmetric shapes are a common feature of 
radio photospheres. Furthermore, secular shape changes were discernible in observations taken several years apart.
This latter finding
suggests that the observed non-spherical shapes most likely result from a combination of pulsation and/or convective effects rather
than rotation or the tidal effects of a companion.

\begin{figure}[t]
  \begin{center}
    \includegraphics[scale=0.3]{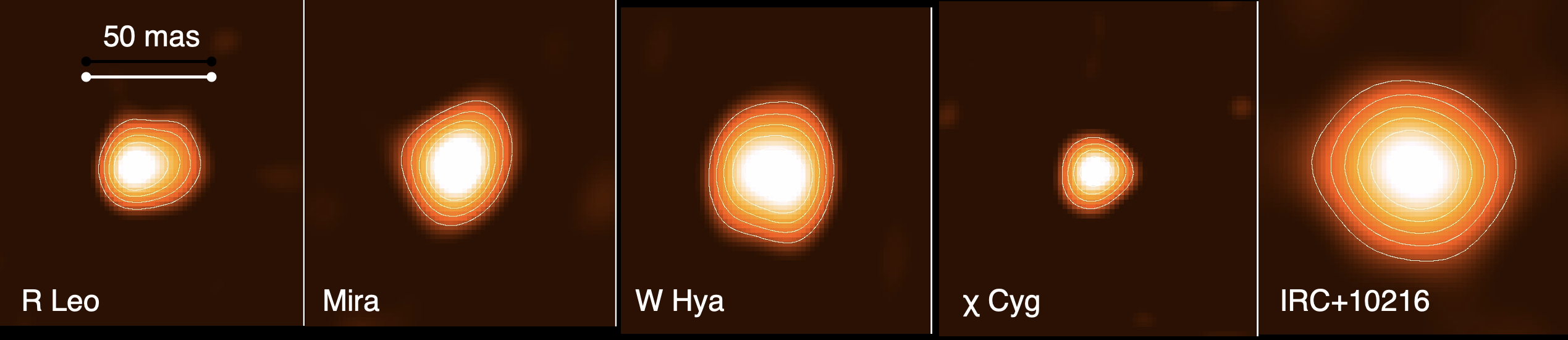}
    \caption{Spatially resolved images of radio photospheres of nearby AGB stars at $\lambda$7~mm, obtained using the Jansky VLA.
      The images were produced using RML imaging techniques, which enabled a modest level of super-resolution
      (see text). Adapted from Matthews \textit{et al.} (2018) and
      Matthews \textit{et al.}, in prep.}
  \label{fig:photospheres}
  \end{center}
\end{figure}

As part of their analysis, Matthews \textit{et al.} (2018) showcased how the interpretation of marginally spatially resolved
observations of radio photospheres
can be further enhanced through the application of a class of radio imaging techniques known  as regularized maximum
likelihood (RML) methods. These imaging algorithms have been exploited recently  to meet the challenges of VLBI imaging at mm
wavelengths using sparse arrays, where traditional {\sc CLEAN} deconvolution tends to perform poorly
(see overview by Fish \textit{et al.} 2016). However, many of the same challenges apply stellar imaging, and
applying RML methods to their $\lambda$7~mm VLA data,
Matthews \textit{et al.} found that it was possible to achieve robust, super-resolved images with resolution as fine as $\sim0.6\times$ the diffraction
limit. This enabled clearly discerning evidence of brightness asymmetries and non-uniformities in the radio photospheres observed with VLA
resolution  which were not visible in {\sc CLEAN} images  (Figure~\ref{fig:photospheres}).
The observed photospheric features appear qualitatively consistent
with the giant convective cells originally predicted to occur in red giant atmospheres
by Schwarzschild  (1975) and that are seen in the bolometric intensity images produced by recent 3D hydrodynamic
simulations (e.g., Figure~\ref{fig:freytag}).
The formation and dissipation of these
cells is suspected of playing an important role in AGB mass loss (e.g., H\"ofner \& Olofsson 2018).

The wavelength dependence of the opacity in radio photospheres (Reid \& Menten 1997)
implies that shorter wavelengths probe successively deeper layers of the
atmosphere. This means that the different wavelength coverages of the VLA and ALMA
are highly complementary for the study of radio photospheres, and that observations of a given star at multiple wavelengths can be used to measure the run of temperature with depth in its atmosphere
(e.g., Matthews \textit{et al.} 2015; Vlemmings \textit{et al.} 2019;
O'Gorman \textit{et al.} 2020). The higher frequencies available at ALMA are also valuable
for providing an additional boost in angular resolution.
While the VLA's 35~km maximum baselines and highest frequency ($\lambda$7~mm receiver) provide a FWHM resolution of $\theta\sim$40~mas
(sufficient to marginally resolve nearby AGB stars within $d\lsim$200~pc),  the combination of ALMA's longest 
baseline configuration (16~km maximum baselines) and Band~7 ($\lambda$0.89~mm) receiver can achieve
$\theta\sim$12--20~mas, sufficient to supply several resolution elements across a nearby AGB star.

Using one such high-resolution ALMA data set at
$\lambda$0.89~mm, Vlemmings \textit{et al.} (2017) reported evidence for a ``hot spot'' on the surface
of the AGB star W~Hya which they interpreted as evidence for a pocket
of chromospheric gas with a brightness temperature $T_{B}>$53,000~K.
The presence of such hot plasma associated with such a cool star ($T_{\rm eff}\approx$2300~K) is confounding, and would seem to require
a combination of strong shock heating and long post-shock cooling times, at odds with current pulsation and convection models. On
the other hand, a re-analysis of the same data by Hoai \textit{et al.} (2022a) seem to show no evidence for the presence of this hot spot on W~Hya,
suggesting the possibility that its origin may have been due to an imaging artifact. Follow-up observations of W~Hya and other similar
stars are clearly of interest to investigate these findings.

\section{Prospects for the Study of AGB Mass Loss with Next Generation Radio Arrays}
Section~\ref{radiophot} described several examples of recent results that illustrate what is possible to achieve
from spatially resolved imaging of the thermal continuum of evolved stars at cm and (sub)mm wavelengths using current state-of-the-art
observational facilities. While these results are both groundbreaking and scientifically valuable, we can anticipate
an enormous leap in such capabilities in the coming decade thanks to planned next-generation radio facilities, including
the
Next Generation Very Large Array (ngVLA; Murphy 2018) and the Square Kilometer Array
(SKA; e.g., Schilizzi 2004; Braun \textit{et al.} 2019).

The ngVLA will be built in the United States and Mexico, and its
``Main Array'' is expected to have $\sim$218 dishes of 18~m diameter spread over
an area several hundred km across. With its combination of frequency coverage (1.2--116~GHz), thermal
sensitivity ($\sim$0.2--0.7~$\mu$Jy beam$^{-1}$ hr$^{-1}$), and
 angular resolution ($\sim$1~mas at 100~GHz), the ngVLA will be a game-changer for stellar imaging (e.g., Figure~6)
 and for the study of evolved stars and their CSEs over all relevant spatial scales, ranging from $\ll R_{\star}$ to $\gsim10^{6}R_{\star}$
 (Matthews \& Claussen 2018; Carilli \textit{et al.} 2018;
 Akiyama \& Matthews 2019).

At the highest angular resolutions of the ngVLA Main Array, some examples of science related to AGB stars and their mass loss that will be enabled  include:

\begin{itemize}
\item The ability resolve radio
surfaces to beyond $d\gsim$1~kpc (thus expanding samples of resolved AGB stars by $\times$300).

\item Resolution of radio surfaces over two decades in frequency for nearby stars ($d\lsim$200 pc).

\item Simultaneous, astrometrically registered studies of photospheric continuum and multiple maser lines.

\item Ability to undertake detailed comparison with (contemporaneous) optical/infrared images from facilities such
  as CHARA and the VLT (see Paladini \textit{et al.} 2018; Ridgway \textit{et al.} 2019).

\end{itemize}
\smallskip

\begin{figure}[t]
  \begin{center}
    \includegraphics[scale=0.45]{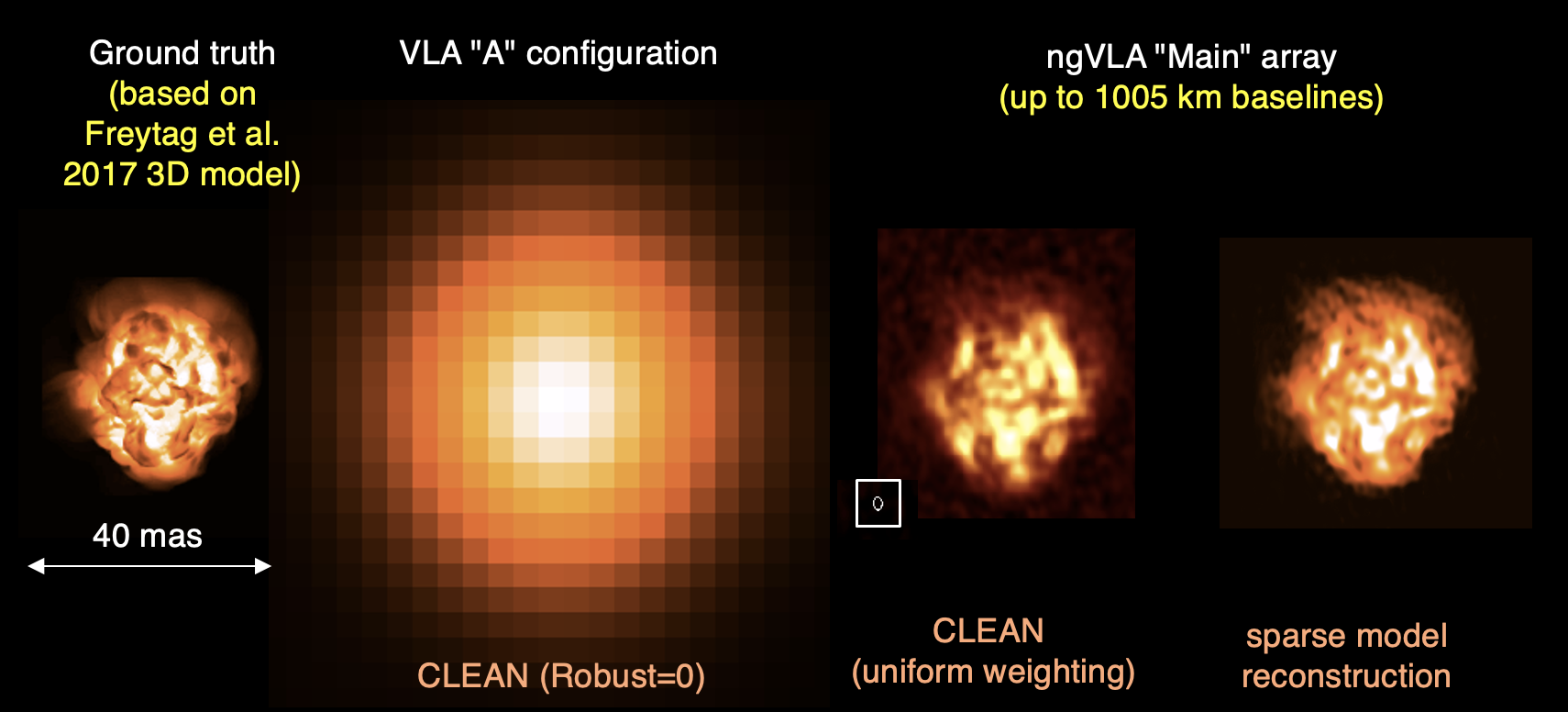}
    \caption{Simulated observations of a radio photosphere at $\lambda$7~mm (46~GHz) for an AGB star at $d$=200~pc.
      A bolometric surface intensity model from Freytag \textit{et al.} (2007; left) was used as a proxy for the expected appearance of
      the thermal radio emission. The
      second image shows a simulated observation of the model with the current VLA `A' configuration (35~km maximum baselines).
      The right two panels
      show 1-hour simulated observations with the ngVLA Main Array (1000~km maximum baselines), imaged using two different
      methods: traditional CLEAN, and a regularized maximum likelihood (sparse modeling) method.
      For additional information, see Akiyama \& Matthews (2019).}
  \label{fig:ngvla}
  \end{center}
\end{figure}

One of the most exciting prospects of the ngVLA for stellar science
will be the ability for the first time, to make ``movies'' of the evolving
radio photospheres of nearby stars over the course of their pulsation cycles  and to quantitatively characterize changes in
 stellar properties over time (Akiyama \& Matthews 2019).
Currently, images of radio photospheres made with the VLA and ALMA have insufficient angular resolution and imaging fidelity to discern subtle
changes in parameters such as stellar radius and brightness temperature with time, or to
chronicle the evolution of surface features that are predicted to occur over timescales of weeks or months (see Figure~\ref{fig:freytag};
see also, e.g., Figure~3 of Freytag \textit{et al.} 2017). However, as shown by Akiyama \& Matthews (2019), this will change
dramatically with the
ngVLA. Indeed, time-lapse movies of the thermal emission should provide exquisite levels of detail comparable to what can now
be seen in  time-lapse movies
of SiO masers with VLBI resolution (cf. Section~6.4.3). Furthermore, it is worth noting that
simultaneous studies of both thermal and maser emission
should provide unprecedented levels of detail for helping to reveal further insights into the mass-loss process of these dynamic and fascinating stars.

The SKA mid-frequency array will be built in the Karoo desert of South Africa, and its initial design (SKA1-Mid) is expected to
cover frequencies from 350~MHz to 15.4~GHz.
Because of its shorter maximum baselines ($\le$150~km) and more limited frequency coverage, 
SKA1-Mid will not be able to rival the ngVLA for spatially resolved stellar imaging, though it will be able to
moderately resolve radio photospheres at $d\lsim$200~pc. In addition, it will be a powerful tool for obtaining sensitive radio
light curves of hundreds of evolved stars. As shown by Menten \& Reid (1997; see also Reid \& Goldston 2002),
the measurement of radio light curves supplies valuable information on the amplitudes of shocks in AGB star atmospheres,
even in the absence of spatially resolved measurements. However, radio light curves are currently available for only a handful
of AGB stars. Obtaining useful  light curves requires a combination of good sensitivity
(typical flux densities are $\lsim$1~mJy in cm bands), accurate calibration (better than $\sim$10\%), and both frequent and long-term temporal sampling
(every 1--2 weeks over timescales of many months). For these reasons, such measurements are technically and logistically
challenging with current arrays. However,
SKA should be able to produce the most accurate radio light curves for AGB stars to date, with quasi-simultaneous coverage
across a wide range of frequencies
(see also Marvel 2004).

\section{Summary}
The mass loss that occurs during the late stages of stellar evolution has implications for a wide range of problems in
astrophysics. However, we do not yet have a complete and fully predictive theory. Sophisticated new 3D hydrodynamic models
for AGB star atmospheres are now available---including pulsation, convection, and other vital physics---that
make highly detailed predictions that can be confronted with observations. However, rigorously testing such models requires
access to observations that spatially resolve the stellar atmosphere and wind launch region on scales ($r\lsim$10~AU), and
temporally resolve relevant dynamical timescales (which can span days to months to years). In this review, I have highlighted
examples of recent cm and (sub)mm wavelength observations, including observations of molecular masers and thermal continuum emission,
that are making progress in these areas and helping to advance our understanding of late-stage stellar mass loss.
Even greater advances are anticipated in the next decade when a new generation of radio telescopes, including the ngVLA and SKA,
comes online.

\ackname{:~~LDM was supported in part by grant AST-2107681 from the National Science Foundation.}


\end{document}